\documentclass[onecolumn,floatfix,12pt]{revtex4}
\pdfoutput=1
\usepackage{graphicx}
\usepackage{dcolumn}
\usepackage{amsmath}
\usepackage{amssymb}
\usepackage{pdflscape}
\usepackage{mathptmx}
\begin{document}

\title [16pt]
{Magic numbers for shape coexistence}

\author
{I. E. Assimakis$^1$, Dennis Bonatsos$^1$, Andriana Martinou$^1$, S.~Sarantopoulou$^1$, S. Peroulis$^2$, T. Mertzimekis$^2$, and N. Minkov$^3$}

\affiliation
{$^1$Institute of Nuclear and Particle Physics, National Centre for Scientific Research 
``Demokritos'', GR-15310 Aghia Paraskevi, Attiki, Greece}

\affiliation
{$^2$ University of Athens, Faculty of Physics, Zografou Campus, GR-15784 Athens, Greece}

\affiliation
{$^3$Institute of Nuclear Research and Nuclear Energy, Bulgarian Academy of Sciences, 72 Tzarigrad Road, 1784 Sofia, Bulgaria}

\begin{abstract}

\parindent=0pt   
\textbf{Abstract:}  The increasing deformation in atomic nuclei leads to the change of the classical magic numbers (2,8,20,28,50,82..) which dictate the arrangement of nucleons in complete shells. The magic numbers  of the three-dimensional harmonic oscillator (2,8,20,40,70...)  emerge at deformations around  $\epsilon=0.6$. At lower deformations the two sets of magic numbers antagonize, leading to shape coexistence.   
 A quantitative investigation is performed using the usual Nilsson model wave functions and the recently introduced proxy-SU(3) scheme.
\end{abstract}

\maketitle

\section{Introduction}

     Shape coexistence \cite{1} is a nuclear phenomenon which is observed when the same nuclei at similar energies can be found in different intrinsic shapes (prolate-oblate, prolate-prolate). Shape coexistence in even nuclei occurs when a ground state band based on the $0_1^+$ ground state is accompanied by a low lying $0^+$ band of  clearly different structure. Good examples are provided by the $_{82}$Pb and $_{80}$Hg isotopes as well as by the $_{50}$Sn isotopes \cite{1}. The existence of the additional band is attributed to two-particle--two-hole (2p-2h) excitations across the proton shell gaps 82 and 50 respectively \cite{1}.
     
     A detailed map of regions of shape coexistence can be found in a recent review article \cite{1}.      The determination of the regions at which shape coexistence appears is an open problem, but a closer look to the  regions shows an underlying dependence  both on the magic numbers of the shell model and the harmonic oscillator.
 
     The purpose of this contribution is the detailed study of the energy gaps that appear with increasing deformation using the Nilsson model \cite{2} with the asymptotic wave functions both for the shell model case and the recently introduced proxy-SU(3) scheme \cite{3}. In the large deformation limit the same set of gaps is obtained in both cases, coinciding with the gaps developing within the three-dimensional harmonic oscillator (3D-HO), since the relative contribution of the matrix elements of the spin-orbit and $l^2$   interactions is decreasing with the increase of the deformation. Shape coexistence is shown to develop within regions defined by some of these gaps.

     Moreover it is shown that at zero deformation the usual magic numbers prevail, while at large deformation the 3D-HO magic numbers dominate, thus it is expected that at intermediate regions the nuclear wave function is a linear combination of these two extremes. The deformations obtained with respect to the two different sets of magic numbers may explain the appearance of shape coexistence.

\section{The Nilsson Hamiltonian for large deformations} 

In this work the calculations have been performed using the Nilsson model Hamiltonian with cylindrical symmetry \cite{2}
\begin{equation}
H=H_{osc} + v_{ls}  \hbar  \omega_0 ({\bf l} \cdot {\bf s}) + v_{ll}   \hbar  \omega_0  ({\bf l}^2 - \langle {\bf l}^2\rangle_N),
\label{Nil}\end{equation} 
where 
\begin{equation}
H_{osc}={ {\bf p}^2 \over 2M} +{1\over 2}  M  \left[\omega_z^2  z^2 + 
\omega^2_\perp  (x^2+y^2)\right].
\end{equation}

The eigenvalues of the terms $H_{osc}$ and $\langle {\bf l}^2\rangle_N$ are given by
\begin{equation}
E_{osc}=\hbar \omega_{0} \left[ N+\frac{3}{2}-\frac{1}{2}\epsilon(3n_{z}-N) \right]
\end{equation}
and
\begin{equation}
\langle {\bf l}^2\rangle_N= {1\over 2} N(N+3)
\end{equation}
respectively.
In the above equations $M$ is the mass of the nucleus, $\bold{s}$  is the spin, $\bold{p}$ is the momentum, while $N$ is the principal oscillator quantum number. The rotational frequencies $ω_z$ and $ω_{\perp}$ are related to the deformation parameter $\epsilon$ by
\begin{equation}
\omega_z = \omega_0 \left( 1 -{2\over 3} \epsilon \right), \qquad  \omega_\perp = \omega_0 \left( 1 +{1\over 3} \epsilon \right).
\end{equation}

The standard values of the constants $v_{ls}$ and $v_{ll}$, can be found in \cite{2}. The terms $\bold{l \cdot s}$ and $\bold{l^2}$ must be diagonalized numerically. This is accomplished by switching from the usual form $K[N n_z \Lambda]$ of  the asymptotic wave functions to the $[n_z r s \Sigma]$ basis. Details of the calculation can be found in \cite{3}.

\section{The proxy-SU(3) scheme} 

     The proxy-SU(3) scheme, which was recently introduced in \cite{3,4}, is an algebraic nuclear model which takes advantage of the SU(3) dynamical symmetry of the 3D-HO. It is known  that the lower shells of nuclei in the shell model have SU(3) as their symmetry group, however as one moves to higher shells the harmonic oscillator structure which is the reason for the SU(3) symmetry is destroyed mostly due to the spin-orbit interaction.
    
     The experimental observation of large spatial overlaps between the orbitals of proton-neutron pairs differing by $\Delta K [\Delta N \Delta n_z  \Delta \Lambda] = 0[110]$ has led to the idea of substituting some orbitals with their 0[110]  counterparts in order to create harmonic oscillator shells \cite{3}.

     More specifically for a given shell, the orbitals of different parity that have invaded the shell from above are replaced with their 0[110] counterparts which were pushed to the shell below. The validity of this approximation was shown in \cite{3} by using a Nilsson calculation as  mentioned above.

     An example of the approximation is the following. We consider the 50-82 major shell consisting of the  3s$_{1/2}$, 2d$_{3/2}$, 2d$_{5/2}$ and  1g$_{7/2}$ orbitals, which are the pieces of the full sdg shell remaining after the spin-orbit force has lowered the 1g$_{9/2}$ orbitals  into the 28-50 nuclear shell.  In addition, it contains the 1h$_{11/2}$  orbitals, lowered into it from the pfh shell, also by the spin-orbit force. The 1g$_{9/2}$ orbital consists of the Nilsson orbitals 1/2[440], 3/2[431], 5/2[422], 7/2[413], 9/2[404],  which are the 0[110] partners of the 1h$_{11/2}$ Nilsson orbitals 1/2[550],  3/2[541], 5/2[532], 7/2[523], 9/2[514], in the same order.  A pair of these 0[110] partners shares exactly the same values of the quantum numbers corresponding to the projections of orbital angular momentum, spin, and total angular momentum. Thus the orbitals in such a pair are expected to exhibit identical behavior as far as properties related to angular momentum projection are concerned.

     One can thus think of replacing all of the 1h$_{11/2}$ orbitals (except the 11/2[505] orbital) in the 50-82 shell by their 1g$_{9/2}$  counterparts. The 1h$_{11/2}$  11/2[505] orbital has been excluded here since it has no partner in the 1g$_{9/2}$ shell.  This is the sole orbit that has to be dropped in this approximation. After these two approximations have been made, we are left with a collection of orbitals which is exactly the same as the full sdg shell of the spherical 3D-HO, which is known to possess a U(15) symmetry, having an SU(3) subalgebra \cite{5}. Therefore we can expect that some of the SU(3) features would appear within the approximate scheme.

%%%%%%%%%%%%%%%%%%%%%%%%%%% FIG. 1  %%%%%%%%%%%%%%%%%%%%%%%%%%%%%%%%%%%%%%%%%%%

\begin{figure*}[htb]

{\includegraphics[width=75mm]{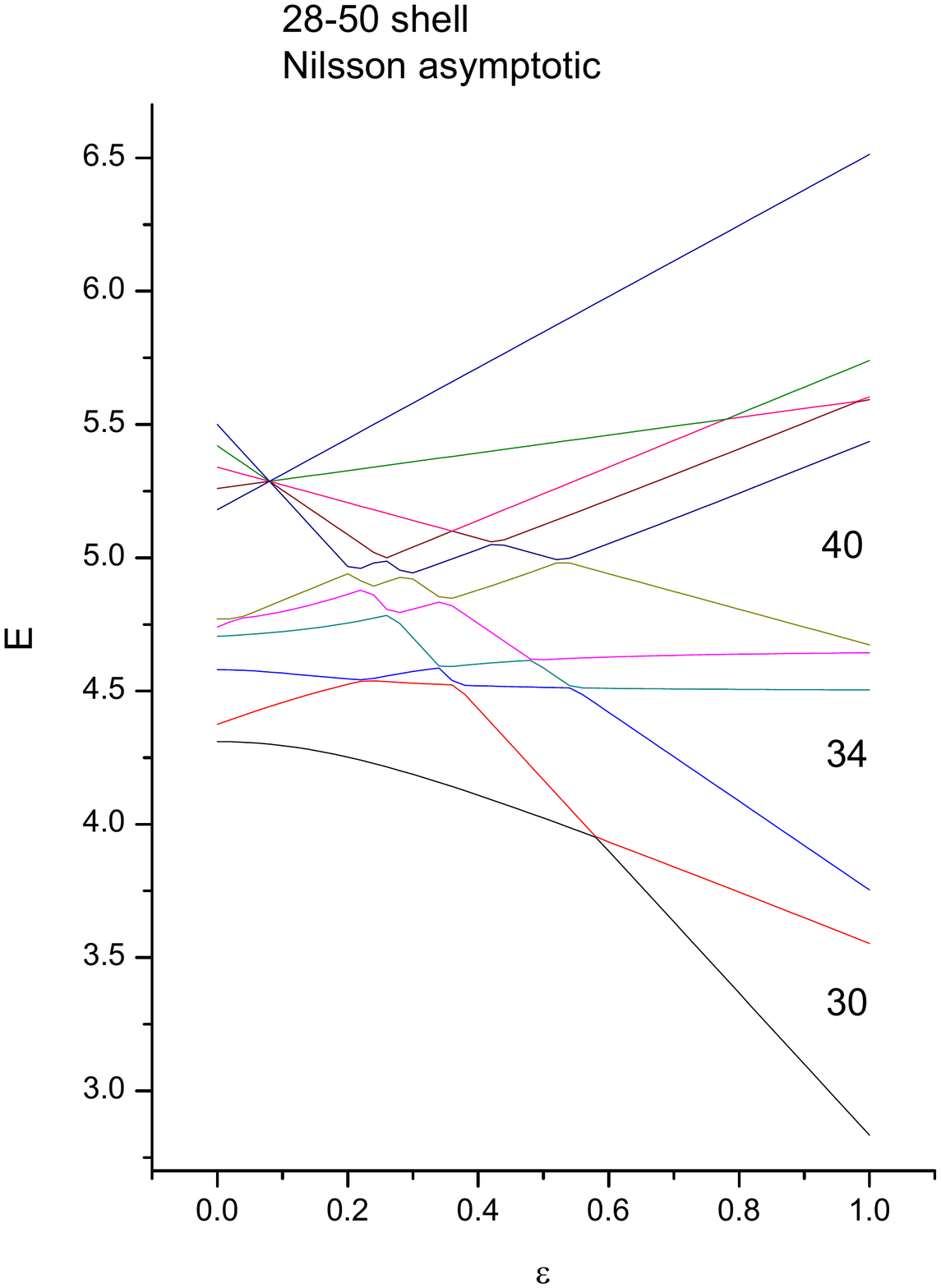}\hspace{5mm}
\includegraphics[width=75mm]{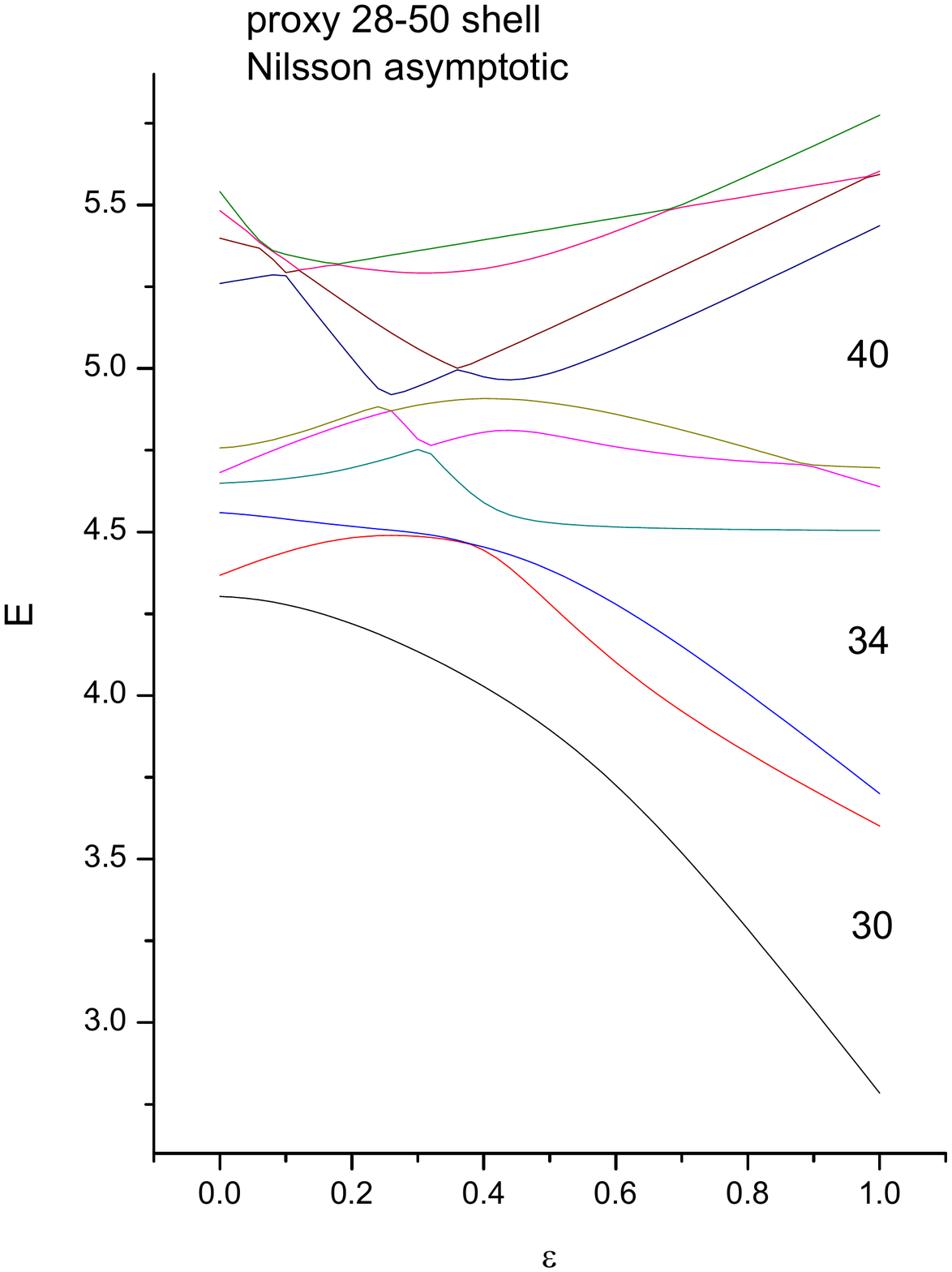}}\vspace{2mm}
{\includegraphics[width=75mm]{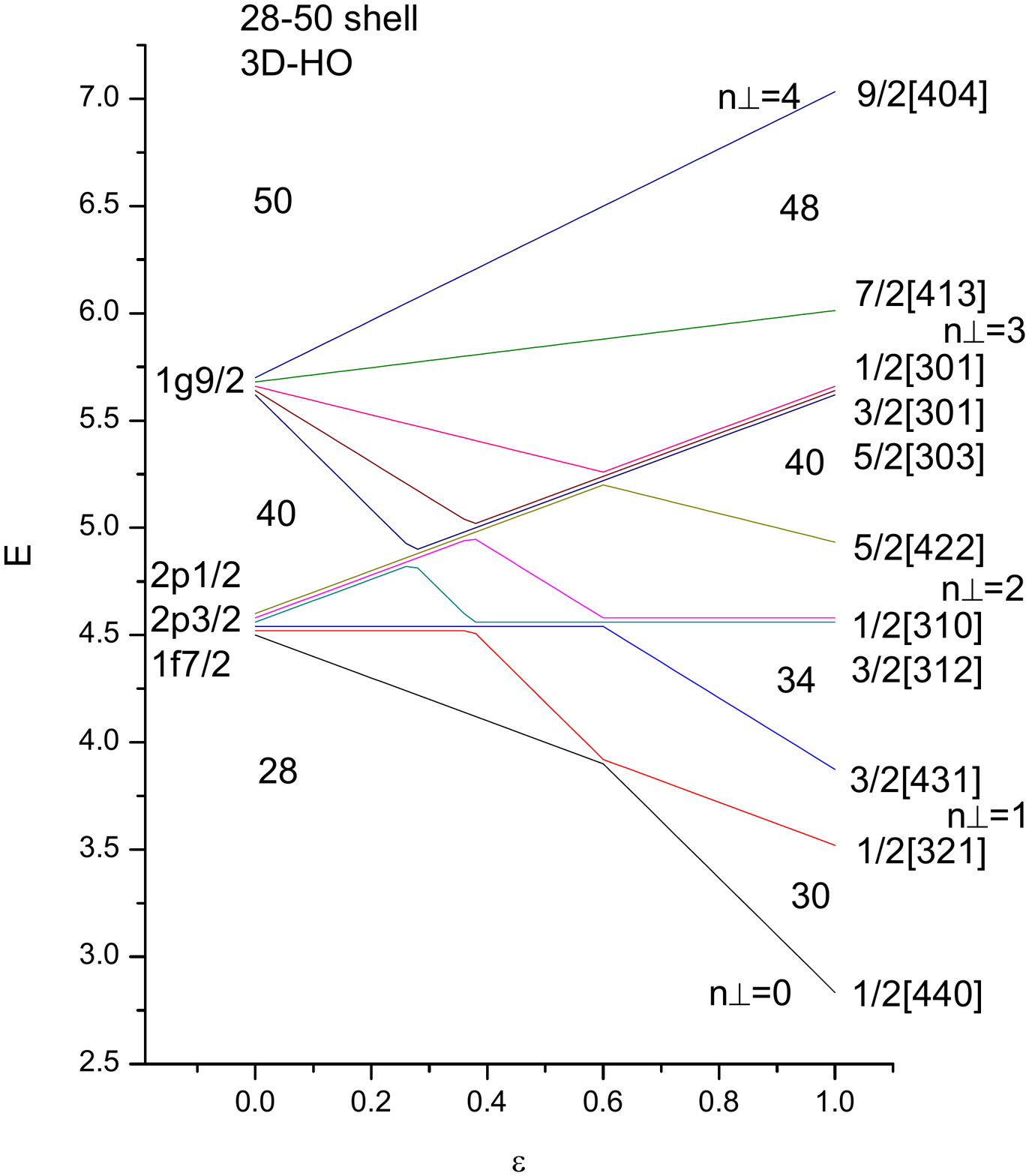}\hspace{5mm}
\includegraphics[width=75mm]{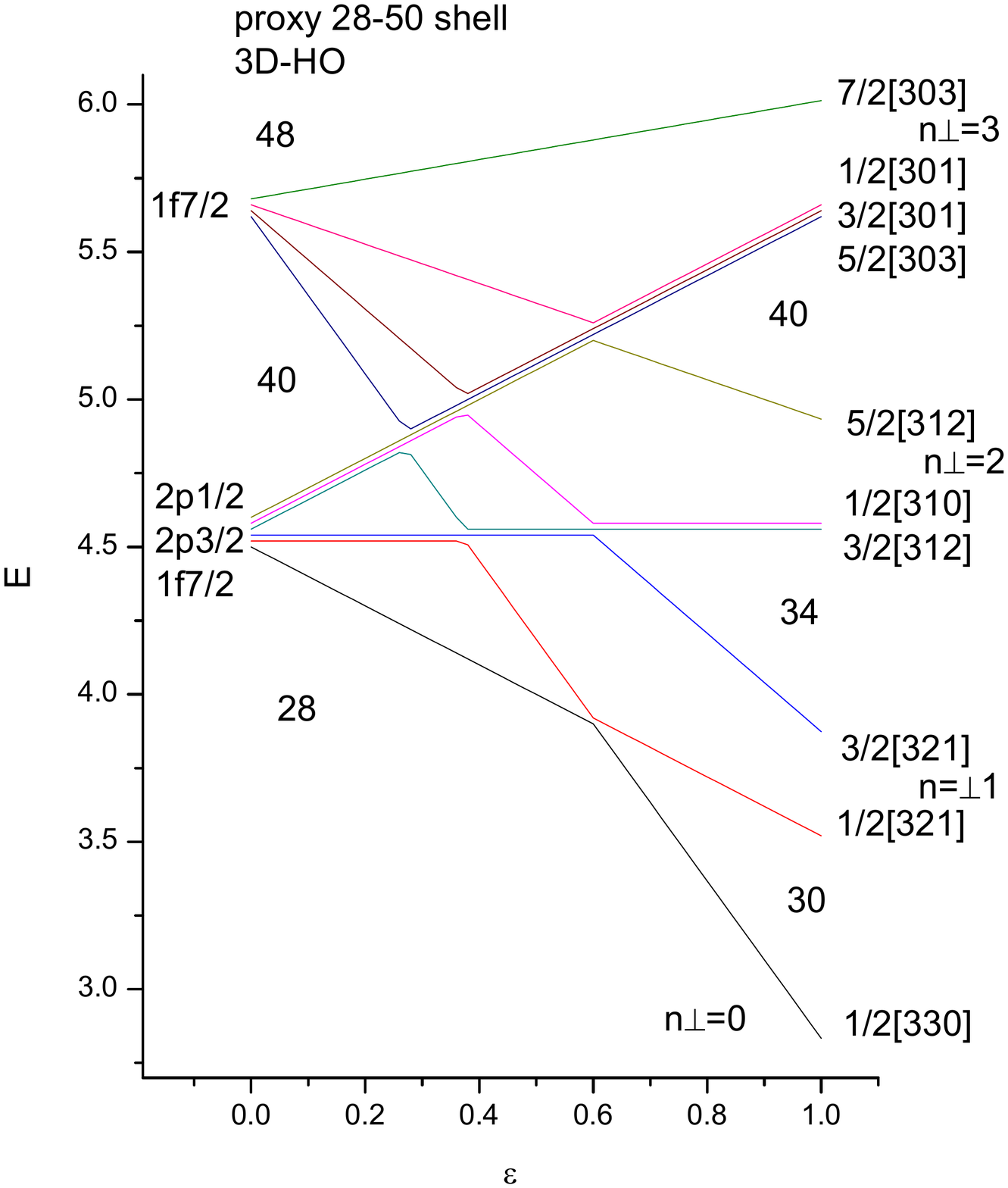}}

\caption{Energies (in units of $\hbar \omega_{0}$) of the Nilsson Hamiltonian as functions of the deformation parameter $\epsilon$ for the 28-50 case.The upper panels correspond to the shell model case.
On the left the usual shell model orbitals have been used, while on the right the orbitals after the 
proxy-SU(3) approximation appear. Results are expected to be valid for $\epsilon>0.15$ due to asymptotic wave functions used. 
}\label{fig1} 

\end{figure*}

%%%%%%%%%%%%%%%%%%%%%%%%%%% FIG. 2  %%%%%%%%%%%%%%%%%%%%%%%%%%%%%%%%%%%%%%%%%%%

\begin{figure*}[htb]

{\includegraphics[width=75mm]{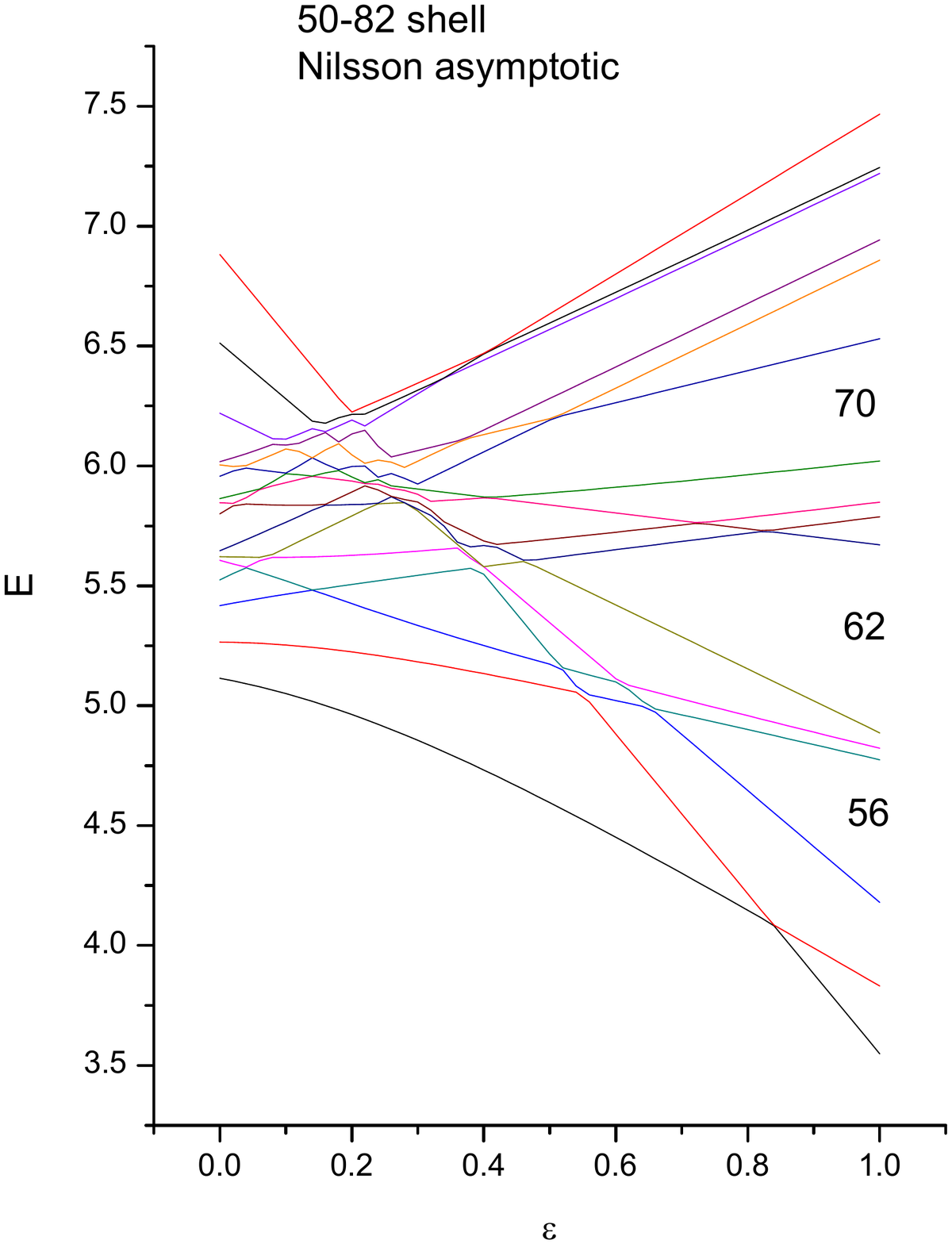}\hspace{5mm}
\includegraphics[width=75mm]{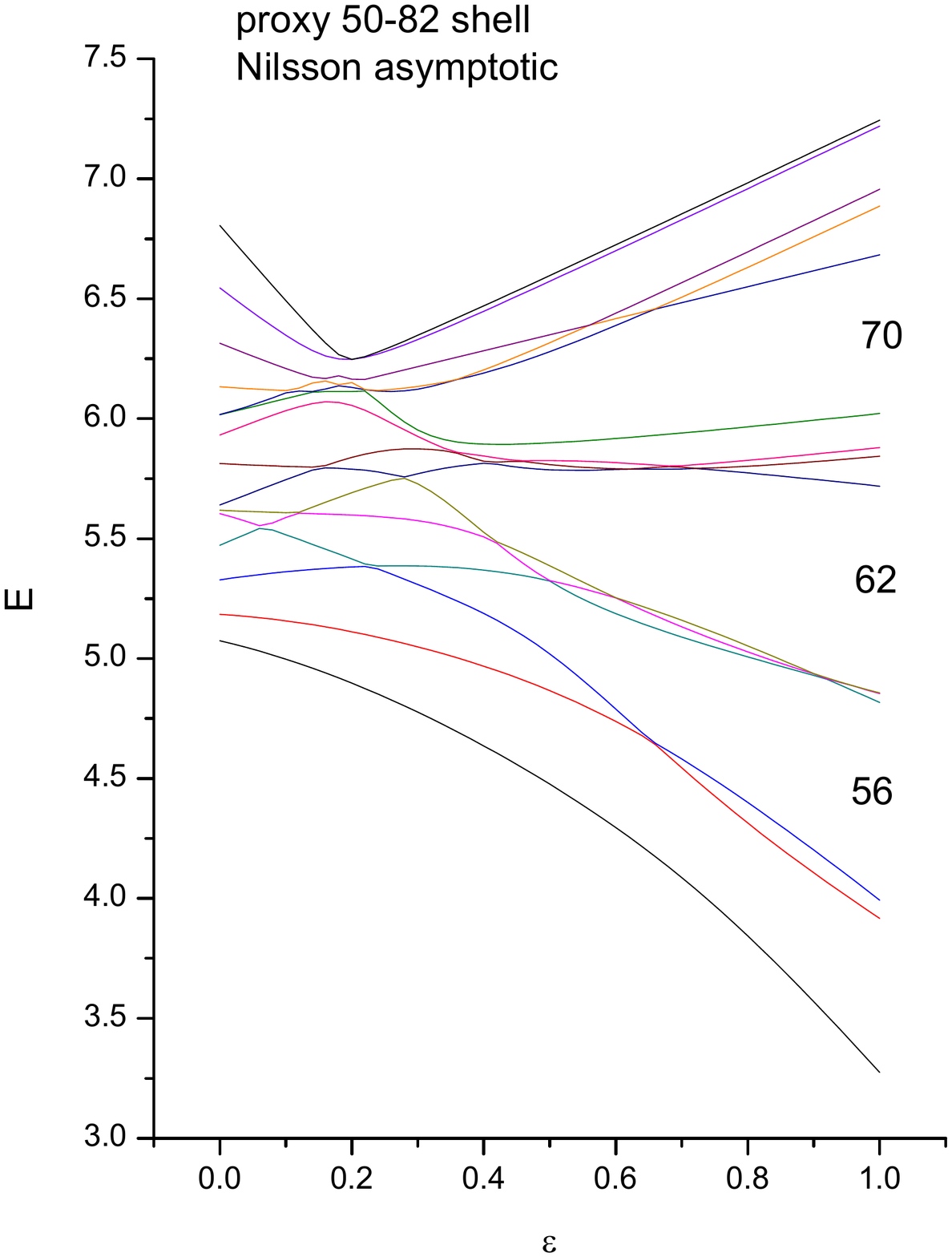}}\vspace{2mm}
{\includegraphics[width=75mm]{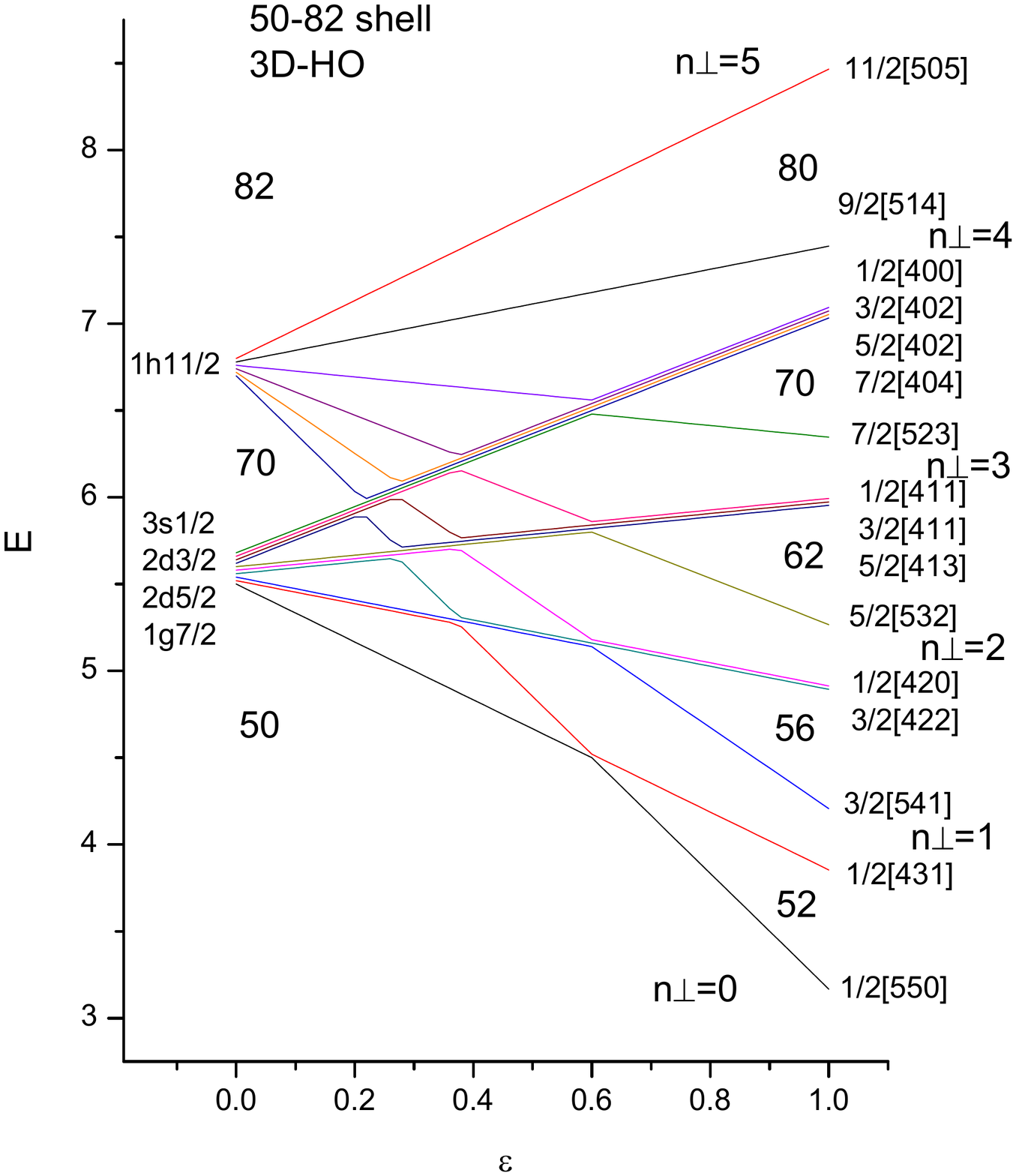}\hspace{5mm}
\includegraphics[width=75mm]{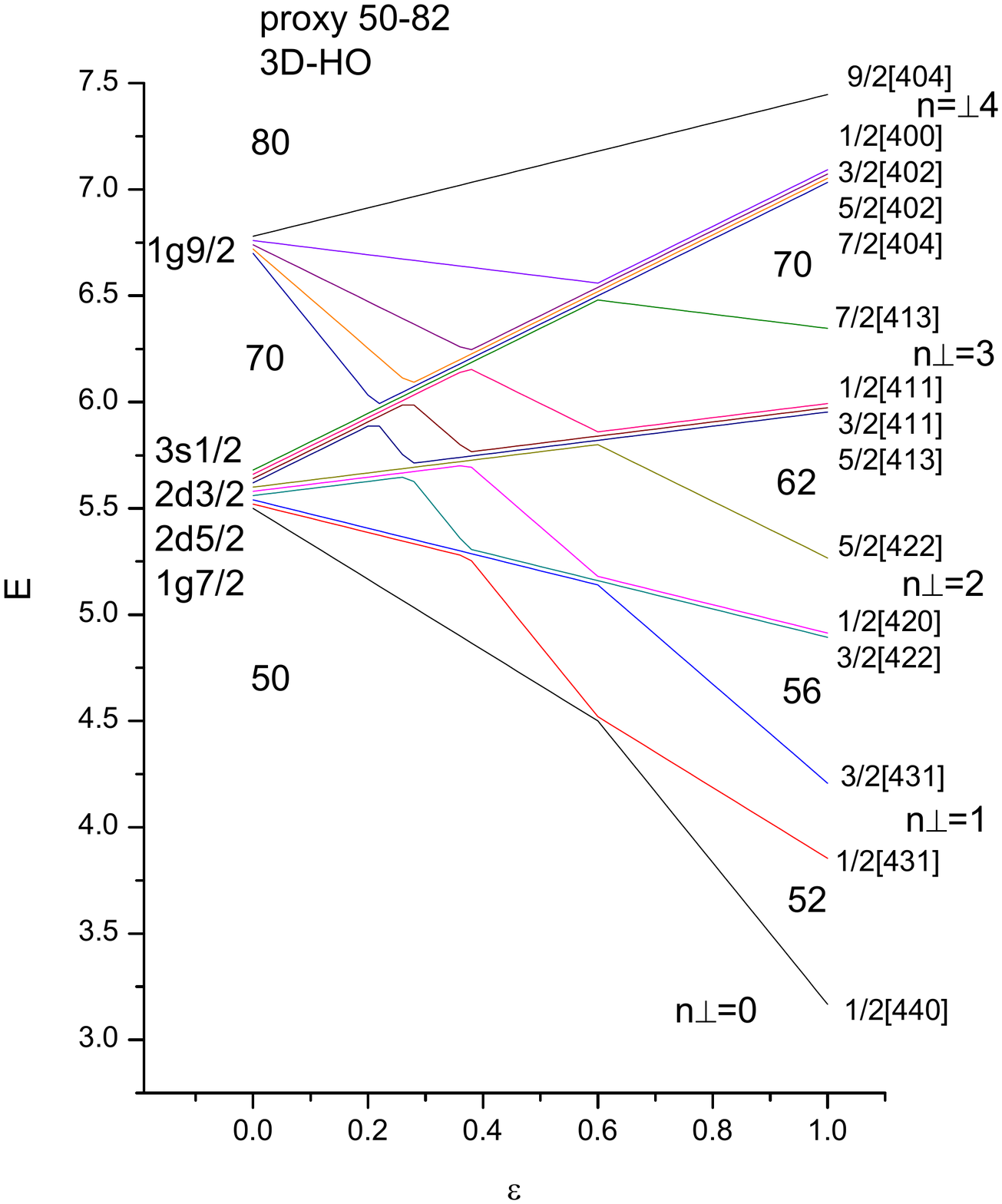}}

\caption{Same as Fig. 1, but for the 50-82 case.
}\label{fig2} 

\end{figure*}

\section{Shell gaps at large deformations}

      The 3D-HO at zero deformation is known to possess the magic numbers 2,8,20,40,70,... However the increase of deformation leads to a change of magic numbers and the next clear set  is obtained [6] at 
      $\epsilon=0.6$~. This corresponds to prolate shapes with axis ratio $\omega_\perp / \omega_z =2∶1$
        and reads  2,4,10,16,28,40,60,80,110,...
    
     In atomic nuclei the magic numbers of the 3D-HO are radically modified by the spin-orbit interaction. At zero deformation the well known magic numbers are 2,8,20,28,50,82... Like  in the harmonic oscillator case, the magic numbers are expected to change with increasing deformation. In what follows we try to examine how the magic numbers change.

     The Nilsson Hamiltonian mentioned above is used alongside with the asymptotic wave functions both for the shell model case and the proxy-SU(3) model. The calculations performed are identical to those of Ref. \cite{3} and are extended up to $\epsilon=1$ for illustrative purposes. In Figs. 1 and 2 the numerical results for the shells 28-50 and 50-82 are shown both for the normal Nilsson model case and the proxy-SU(3) scheme. The top panels in each figure correspond to the usual shell model case including the spin-orbit interaction, while the lower panels correspond to the pure 3D-HO case without the presence of the spin-orbit interaction. It is obvious that at large deformations the same energy gaps between orbitals appear for both cases. Same conclusions can be made for the proxy-SU(3) case, shown on the rhs panels of the figures, since it has been shown in \cite{3} that proxy-SU(3) is a good approximation. 

     In more detail, in Fig. 1 the results for the 28-50 case are presented. The upper left panel shows the results for the usual shell model case in the Nilsson framework. With increasing deformation shell gaps appear for 30, 34 and 40. The same gaps appear in the lower left panel, where the calculations are performed for the 3D-HO case excluding the spin orbit interaction. Note that 40 is the magic number for the 3D-HO at zero deformation. The same conclusions can be drawn using the proxy-SU(3) model as it is shown at 
     the upper right and lower right panels. The same line of thought is followed in Fig. 2, but for the 50-82 case. Similar conclusions can be drawn. The gaps appearing now are 56, 62 and 70, with 70 being the 3D-HO magic number at zero deformation.

     In Fig. 3  the size of the gaps appearing at 40, 70 in the 3D-HO case are compared with the gaps observed at 50, 82 in the usual shell model case. It can be seen that at large deformations the gaps 40, 70 prevail, while at low deformations both sets of gaps are of comparable size. The same comparisons are made in Fig. 4 with the use of the proxy-SU(3) model. Similar conclusions can be drawn.

     The comparable size of the gaps at intermediate deformations leads to an explanation of the empirically observed shape coexistence. If a specific nucleus has an intermediate value of the deformation parameter 
     $\epsilon$, then its valence nucleons (depending on the case, proxy-SU(3) or 3D-HO) can be counted with respect to different sets of magic numbers. This leads to different descriptions/predictions of the shape of the nucleus as shown  in \cite{4}.

%%%%%%%%%%%%%%%%%%%%%%%%%%% FIG. 3  %%%%%%%%%%%%%%%%%%%%%%%%%%%%%%%%%%%%%%%%%%%

\begin{figure*}[htb]

{\includegraphics[width=75mm]{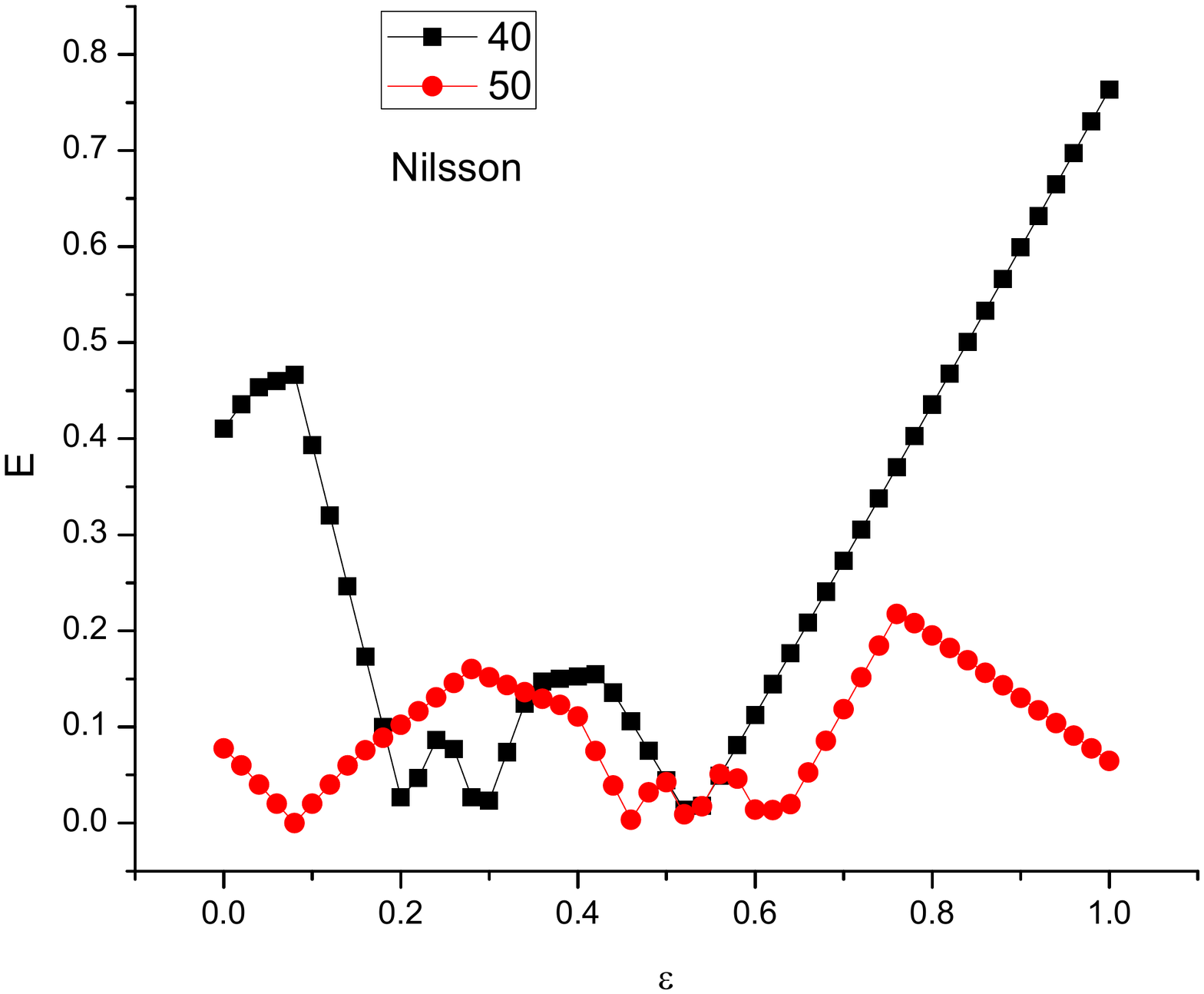}\hspace{5mm}
\includegraphics[width=75mm]{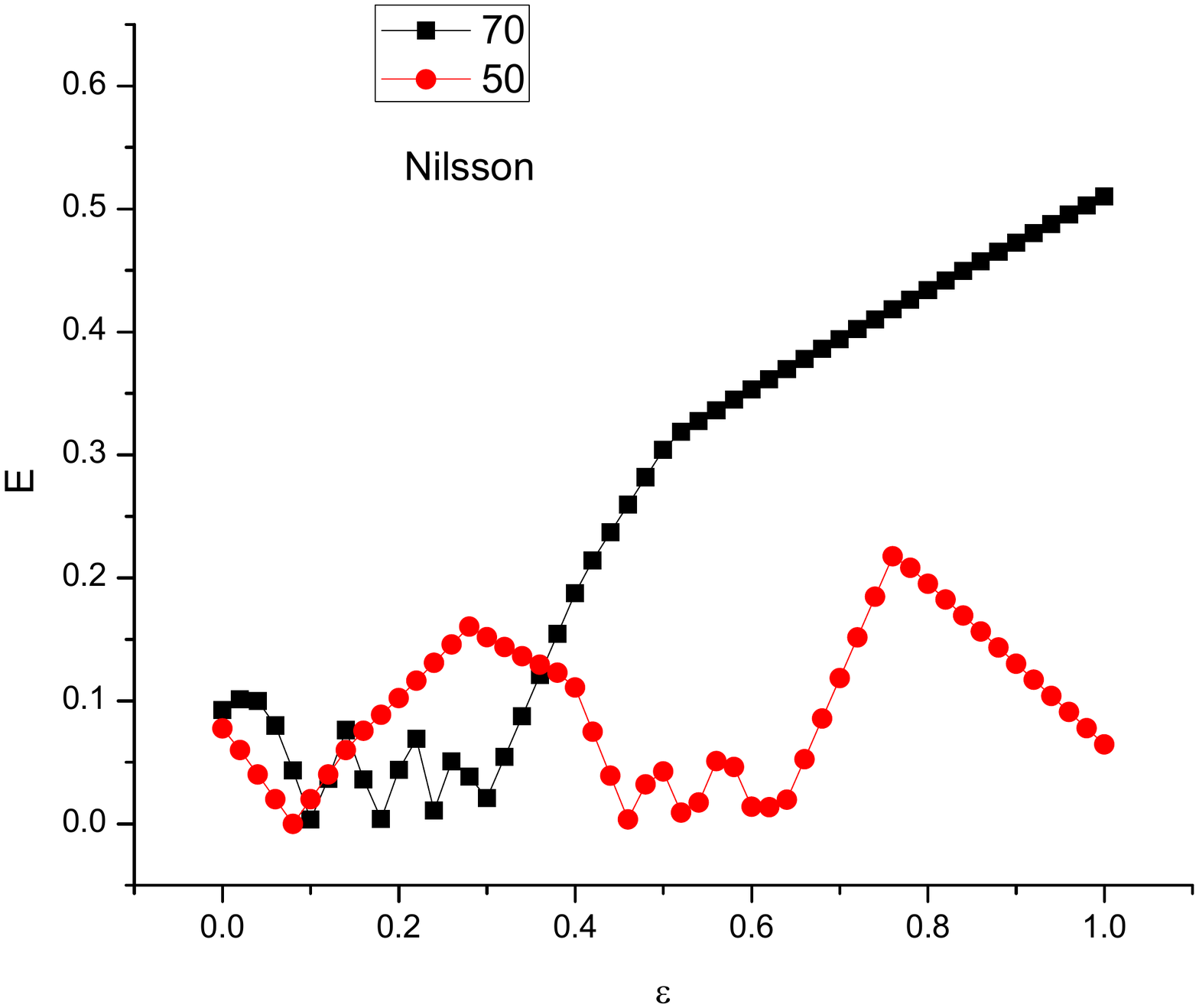}}\vspace{2mm}
{\includegraphics[width=75mm]{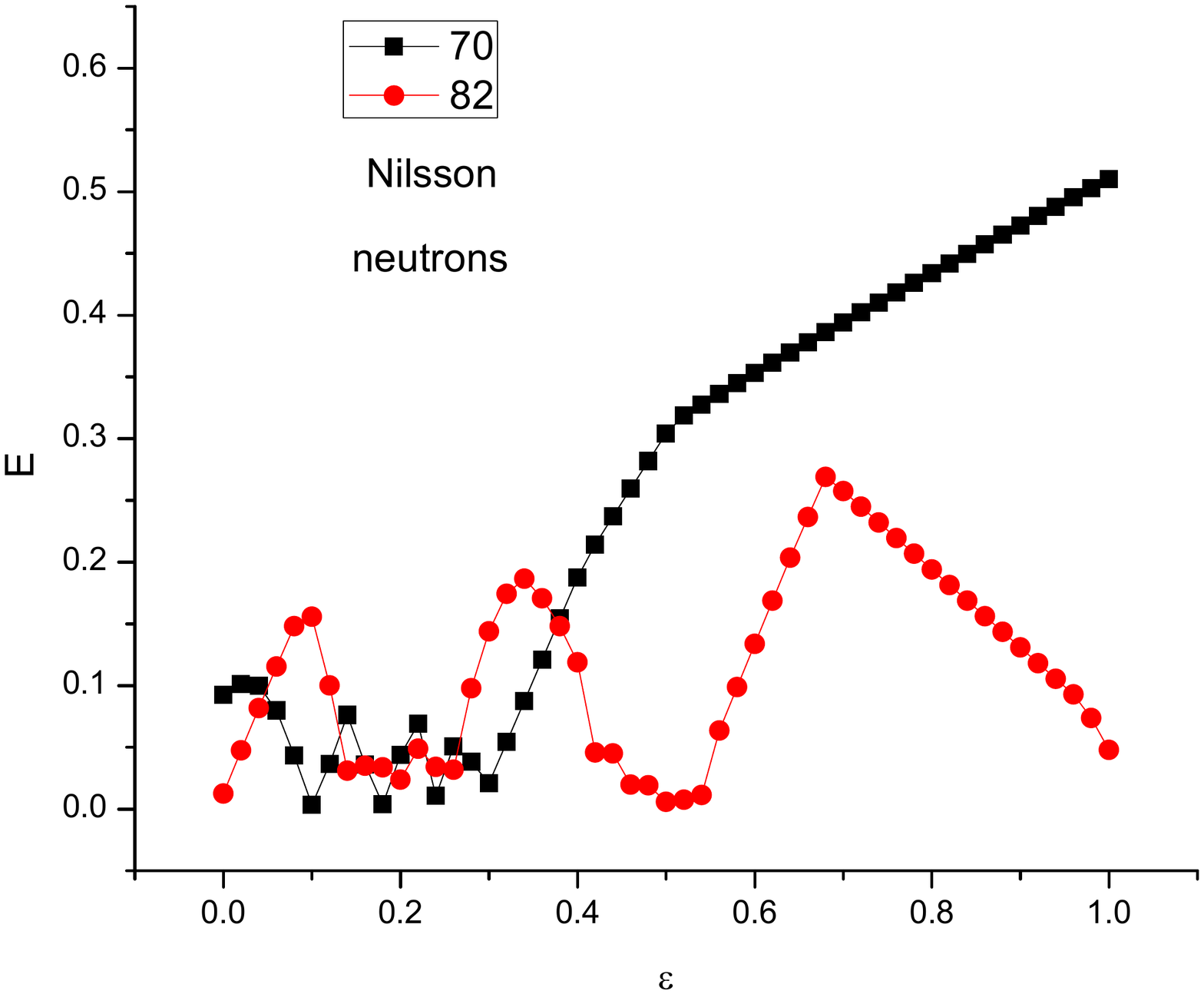}\hspace{5mm}
\includegraphics[width=75mm]{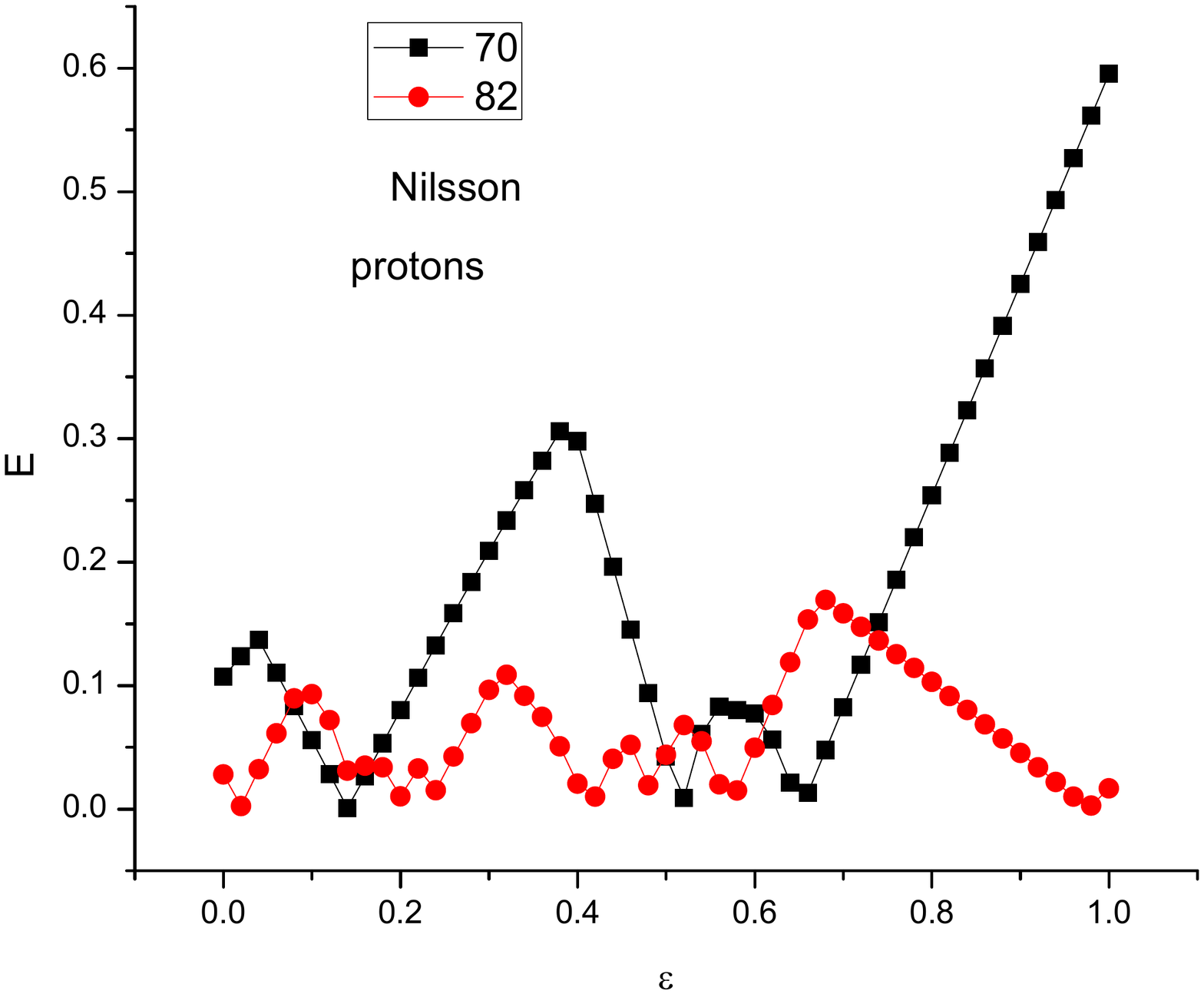}}

\caption{Energy gaps (in units of $\hbar \omega_{0}$) appearing in Figs. 1 and 2 above the nucleon numbers indicated. Due to the use of asymptotic wave functions, results are expected to be valid for $\epsilon>0.15$.
}\label{fig3} 

\end{figure*}

%%%%%%%%%%%%%%%%%%%%%%%%%%% FIG. 4  %%%%%%%%%%%%%%%%%%%%%%%%%%%%%%%%%%%%%%%%%%%

\begin{figure*}[htb]

{\includegraphics[width=75mm]{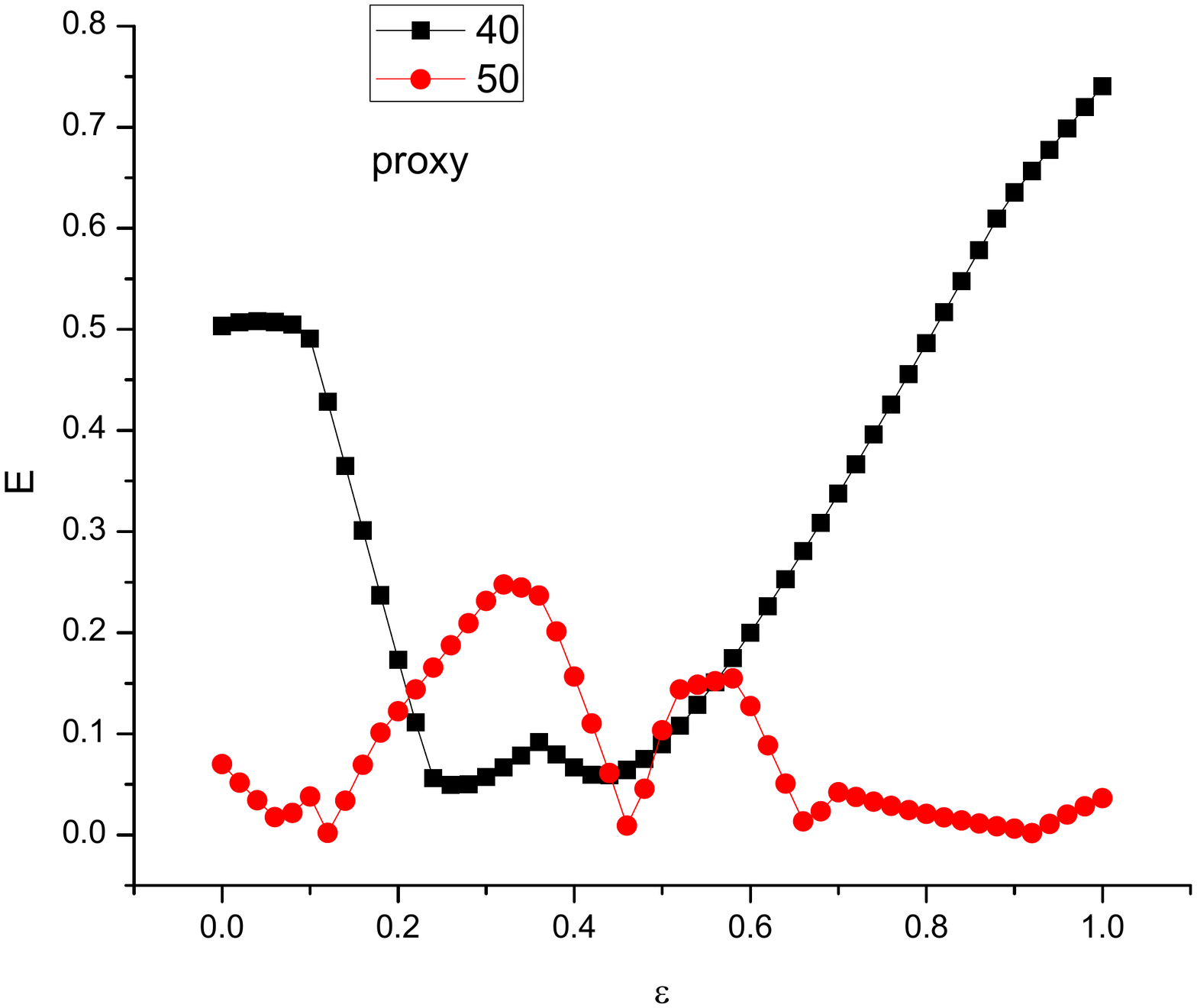}\hspace{5mm}
\includegraphics[width=75mm]{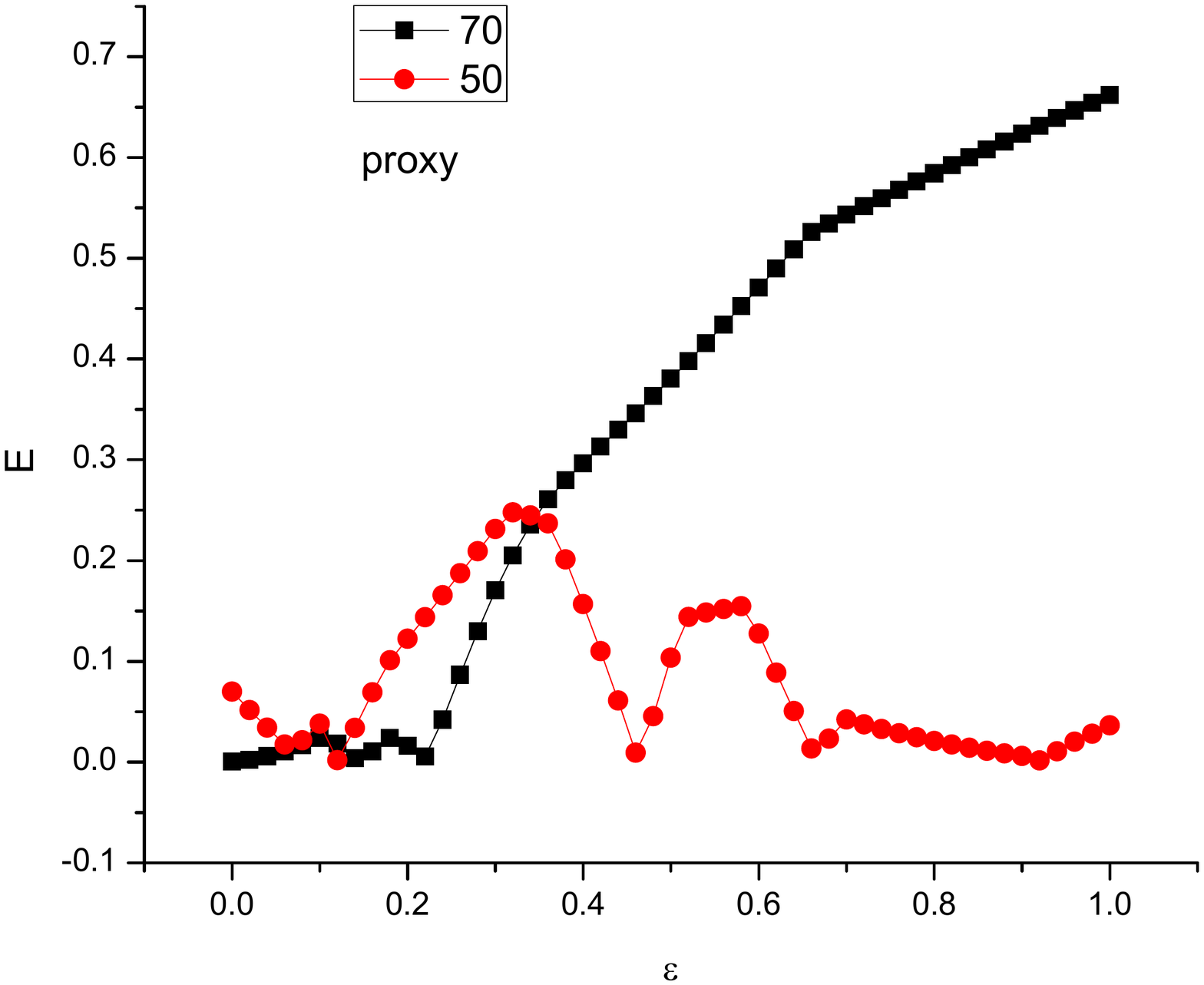}}\vspace{2mm}
{\includegraphics[width=75mm]{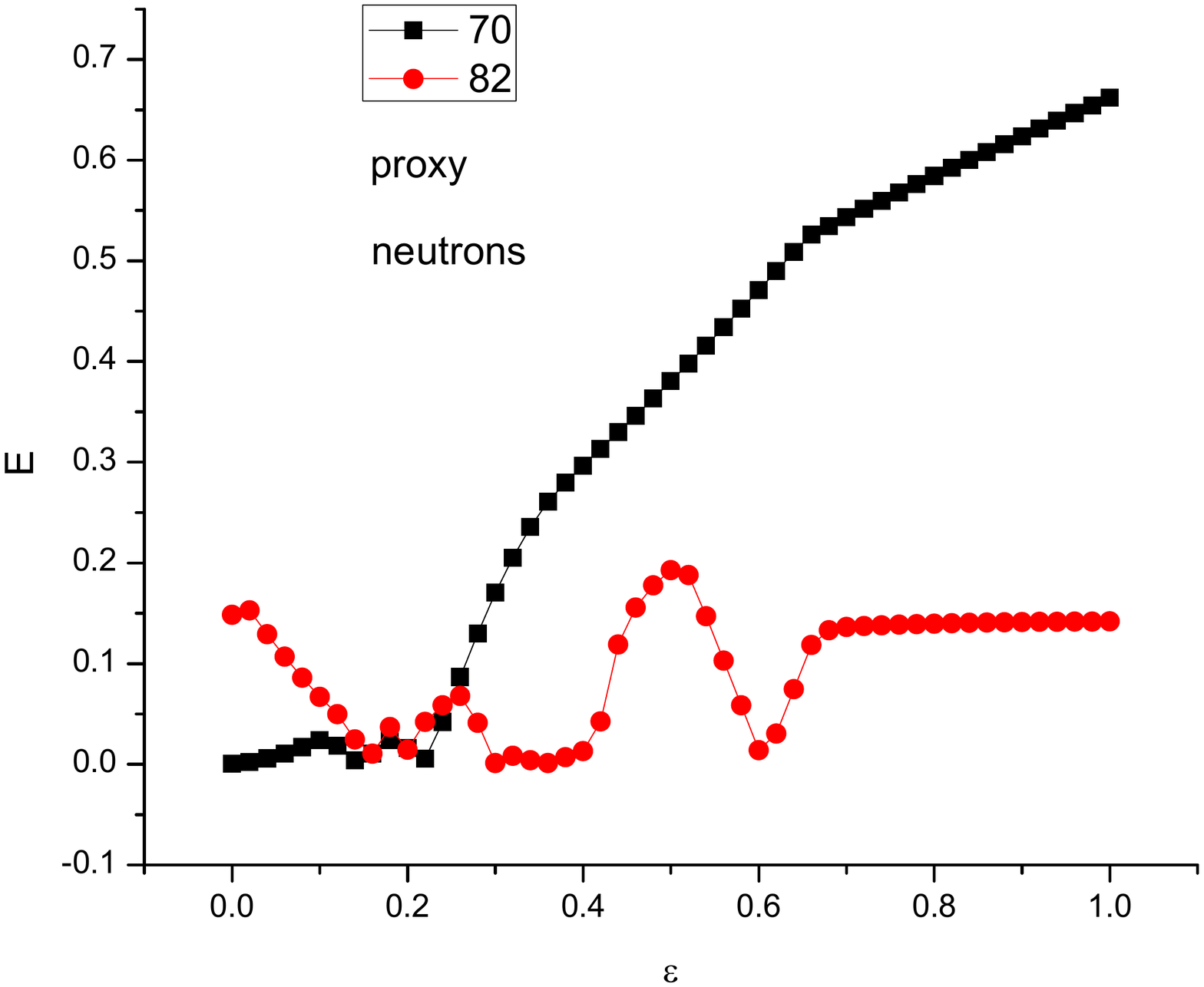}\hspace{5mm}
\includegraphics[width=75mm]{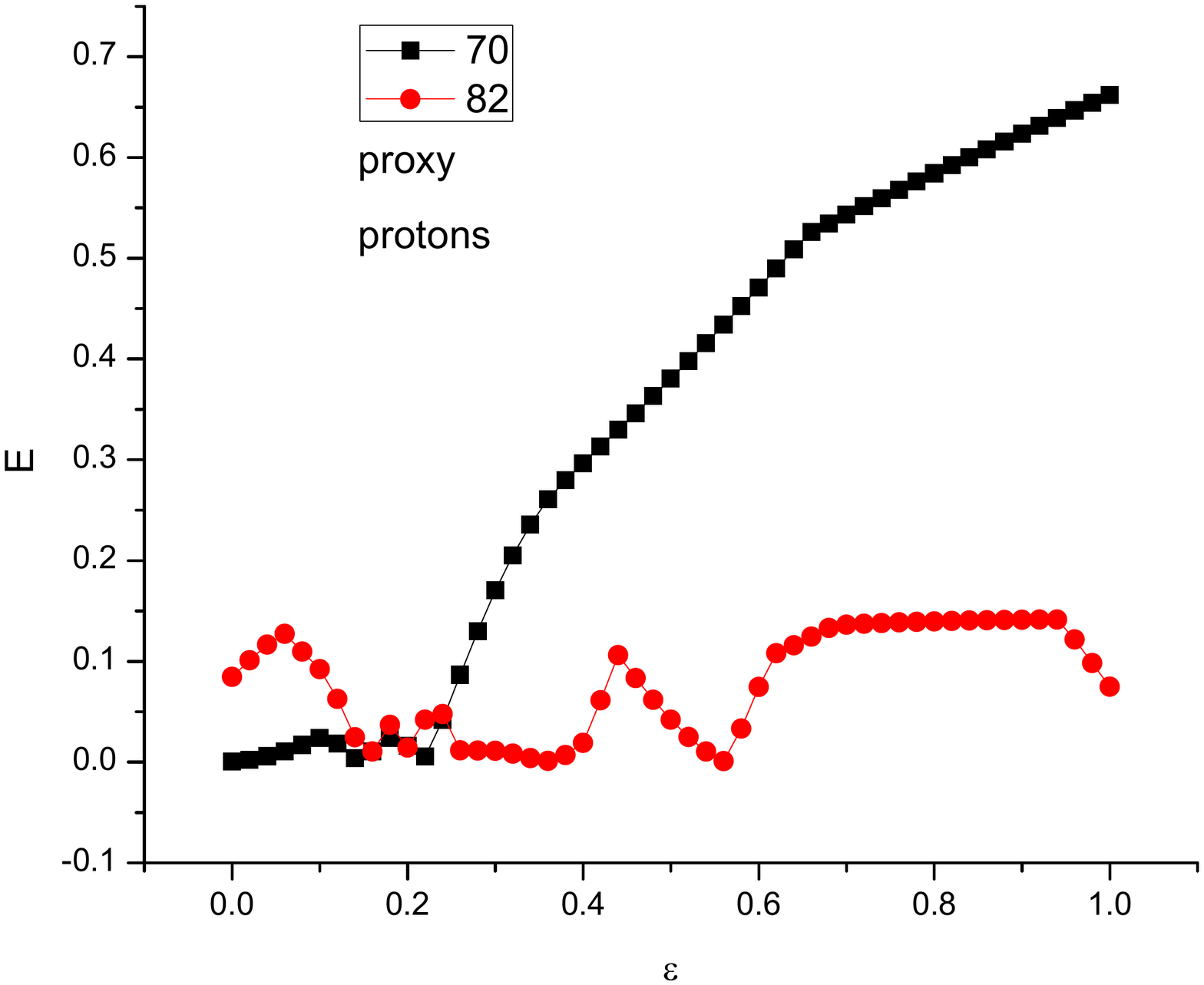}}

\caption{Same as Fig. 3, but for the proxy-SU(3) calculations.
}\label{fig4} 

\end{figure*}

\section{Conclusions} 

With the increase of the deformation, the magic numbers change. At zero deformation the usual nuclear magic numbers prevail while as $\epsilon$ increases the 3D-HO magic numbers begin to emerge. The present findings are consistent with the occurrence of the 3D-HO magic numbers at $\epsilon=0.6$ shown by Sugawara-Tanabe {\it et al.} \cite{6}. Predictions for the regions of shape coexistence are given in \cite{7}. Results for the 82-126 case and comparisons of other sets of magic numbers will be presented in upcoming work .

\end{document}